\pdfoutput=1 

\documentclass[conference]{IEEEtran}
\usepackage{flushend}

\usepackage{footnote}

\usepackage{listings}
\usepackage[cmex10]{amsmath}
\usepackage{bm}

\usepackage[pdftex]{graphicx}
\graphicspath{{./fig/pdf/}{./fig/jpeg/}{./fig/png/}}
\DeclareGraphicsExtensions{.pdf,.jpeg,.png}

\usepackage{cite}

\ifCLASSINFOpdf
\else
\fi
\newcommand{\bmm}{\mathbf{m}}

\newcommand{\bA}{\mathbf{A}}

\newcommand{\bV}{\mathbf{V}}
\newcommand{\bW}{\mathbf{W}}

\begin{document}
%






\title{Signal Processor based on Gaussian Message Passing}

\title{A Signal Processor for Gaussian Message Passing}



%
\author{ \IEEEauthorblockN{Harald Kr\"oll\IEEEauthorrefmark{1}, 
    Stefan Zwicky\IEEEauthorrefmark{1}, 
    Reto Odermatt\IEEEauthorrefmark{1}, 
    Lukas Bruderer\IEEEauthorrefmark{2}, 
    Andreas Burg\IEEEauthorrefmark{3}, 
    Qiuting Huang\IEEEauthorrefmark{1}} 
  \IEEEauthorblockA{\IEEEauthorrefmark{1}Integrated Systems Laboratory, ETH Zurich}
  \IEEEauthorblockA{\IEEEauthorrefmark{2}Signal and Information Processing Laboratory, ETH Zurich}
  \IEEEauthorblockA{\IEEEauthorrefmark{3}Telecommunication Circuits Laboratory, EPF Lausanne}
    Email: \{kroell,zwicky,huang\}@iis.ee.ethz.ch, bruderer@isi.ee.ethz.ch, andreas.burg@epfl.ch}


\maketitle

\begin{abstract}

    %

    %
    %
    %
    In this paper, we present a novel signal processing unit built
    upon the theory of factor graphs, which is able to address a wide
    range of signal processing algorithms.
    More specifically, the demonstrated factor graph processor (FGP)
    is tailored to Gaussian message passing algorithms.
    We show how to use a highly configurable systolic array to solve
    the message update equations of nodes in a factor graph
    efficiently.
    A proper instruction set and compilation procedure is presented.
    In a recursive least squares channel estimation example we show
    that the FGP can compute a message update faster than a
    state-of-the-art DSP.
    The results demonstrate the usabilty of the FGP architecture as a
    flexible HW accelerator for signal-processing and communication
    systems.

\end{abstract}


%
\IEEEpeerreviewmaketitle

\section{Introduction}

%
%
%
Computationally demanding tasks such as channel estimation or
equalization are omnipresent in modern communication and
signal-processing systems.
Algorithms for such tasks might be implemented on hardware
accelerators, on a programmable processor such as a DSP or on an ASIP
with a custom instruction set.
Various attempts have been reported to bridge the gap between
hardwired accelerators and programmable processors, where the
trade-offs between performance and programmability have been subject
of many studies\cite{bridging}.
%
%
%
%
%
%
The goal of this work is to design a signal processor or accelerator
with an appropriate instruction set, which offers sufficient
flexibility to perform a wide range of common signal-processing tasks.
To achieve this, we leverage the theoretical framework of factor
graphs (FGs) \cite{kschischang} and message-passing algorithms.
%
%
More specifically, we focus on FGs that represent factorizations of
(scaled) multivariate Gaussian probability distributions.
%

%
%
%



In~\cite{loel} it is shown how widely-used algorithms such as recursive
least squares (RLS), linear MMSE equalization, and Kalman filtering
can be expressed with Gaussian message-passing (GMP) on a FG.
GMP algorithms have applications in various different fields such as
wireless communications~\cite{Kirkelund}, optical
imaging~\cite{PhaseImg}, Time of Arrival (ToA) estimation~\cite{TOA},
multiuser detection~\cite{CDMA} or compressed sensing~\cite{CS}.

\begin{figure}[ht]
\centering
\includegraphics[width=\columnwidth]{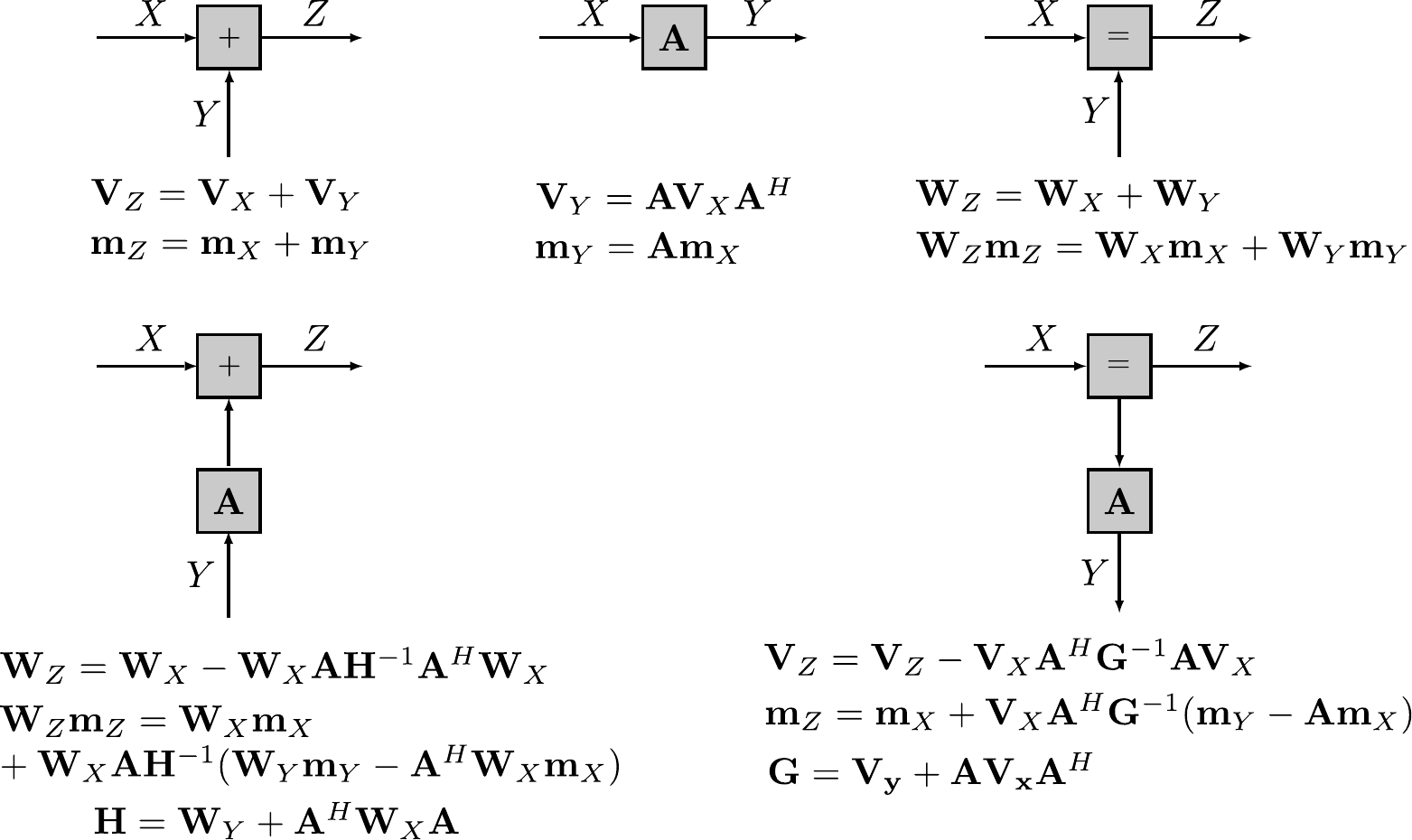}
\caption{FG nodes and the corresponding message
  update rules for GMP.~\cite{loel}}
\label{fig:nodesall}
\end{figure}

A FG is an undirected graph, where each node represents a function.
In GMP messages, which are represented by a mean vector $\bmm$ and a
covariance matrix $\bV$, or a transformed mean vector $\bW\bmm$ and a
weight matrix $\bW$ are exchanged between nodes of a graph.
The computational load is distributed among the nodes, which compute
updates of the outgoing messages based on the incoming messages.
The various types of nodes (i.e., functions) that are used in GMP are
summarized in Fig.\,\ref{fig:nodesall}. The figure also displays the
message update rules corresponding to each type of node.

To the best knowledge of the authors, there exists no programmable
processing architecture with a datapath suited for solving GMP
algorithms.
%
%

In this paper, we present a novel application specific instruction
processor termed FGP which is able to perform the message updates for
all basic types of nodes in GMP. Therefore, complex algorithms
described by GMP can be readily executed on our processor. The
computations in the FGP are based on a systolic array implementation
of the Faddeev algorithm, which is able to efficiently compute all
necessary matrix operations. Eventually, we demonstrate an appropriate
compiler for the FGP and compare the throughput of the proposed
implementation with a state-of-the-art DSP for a compound node message
update.

\section{From GMP nodes to datapath architecture}
\label{sec:datapath}

\begin{figure}[ht]
\centering
\includegraphics[width=\columnwidth]{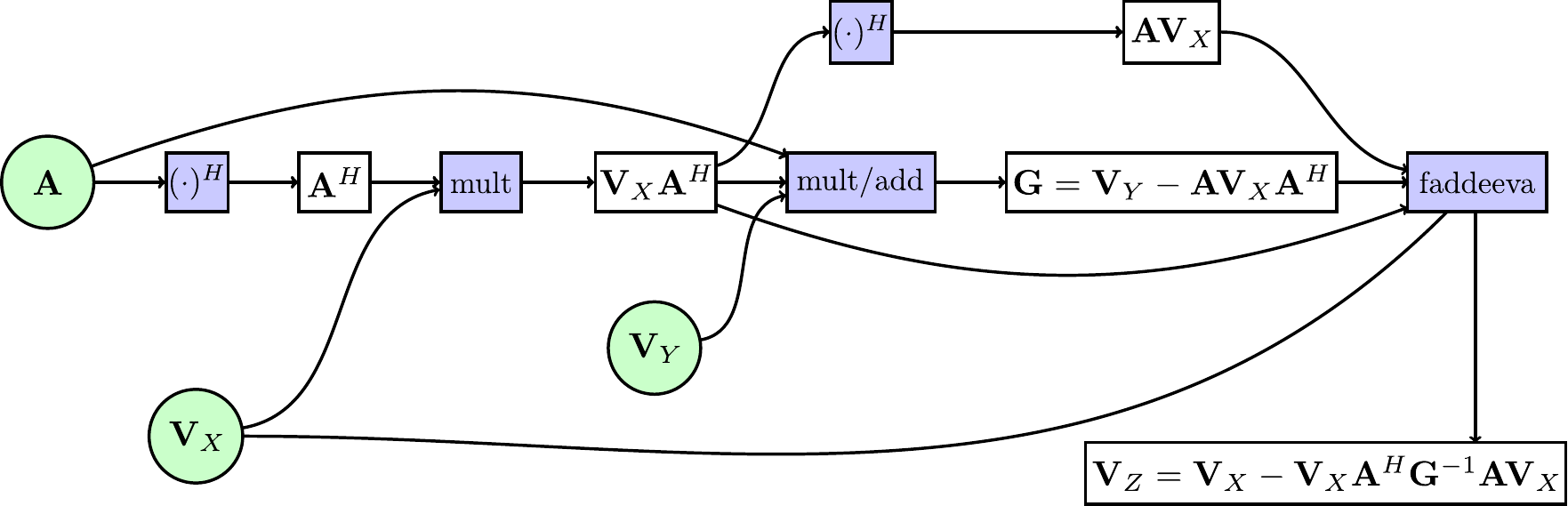}
\caption{Data dependency graph of operations in compound node.}
\label{fig:cn}
\end{figure}

The FGP architecture is derived from the message update rules of
the nodes in Fig.\,\ref{fig:nodesall}.
Compound nodes are composed by two simple nodes.
The expressions for the compound nodes consist of the same basic
matrix operations as the ones for the simple nodes what facilitates
the development of a datapath which is suitable for all nodes.
The data dependency graph for the message update (covariance matrix
only) computation of a compound node is shown in Fig.~\ref{fig:cn}.

The outgoing message $\mathbf{V}_Z$ is computed with the incoming
messages $\mathbf{V}_X$, $\mathbf{V}_Y$ and a state matrix $\bA$
defined the particular node.
The purple boxes highlight the computations to be carried out, while the
white boxes contain the intermediate results.
The computations can be decomposed into three types\footnote{Hermitian
  transp. and negation are not considered as a separate operations.}:

\begin{itemize}
\item Matrix-matrix multiplication (e.g., $\bV_X \bA^H$)
\item Matrix-matrix multiplication with addition/subtraction (e.g., $\bV_Y - \bA (\bV_X \bA^H)$)
\item Schur complement (e.g., $\mathbf{V}_X - (\mathbf{V}_X \mathbf{A}^H) \mathbf{G}^{-1} (\mathbf{AV}_X)$)
\end{itemize}

\begin{figure}[h!]
\centering
\includegraphics[width=\columnwidth]{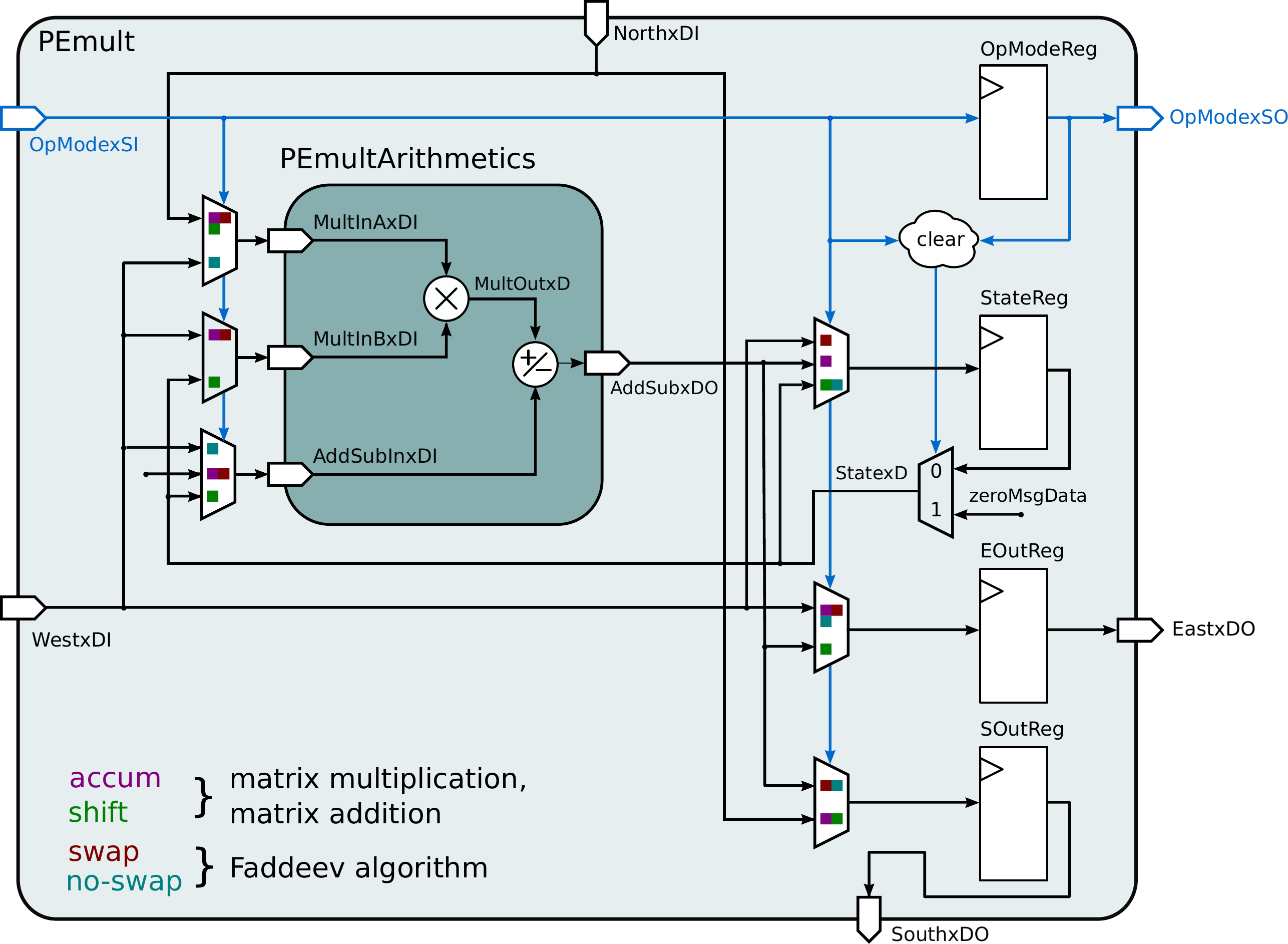}
\caption{{\em PEmult} with different operation modes.}
\label{fig:PEcenter}
\end{figure}

%
A suitable architecture for matrix computations of such a kind
is the systolic array~\cite{syst_kalman}.
%
%
%
%
For the matrix-matrix multiplication ($\bV_X \bA^H$), a
two-dimensional rectangular systolic array as shown in the blue box in
Fig.\,\ref{fig:FGP} is sufficient.
The processing elements (PEs) need to perform multiplications and
additions.
This so-called \textit{PEmult} is shown in Fig.\,\ref{fig:PEcenter}.

It supports different operation modes which will be explained next and
contains both a real-valued multiplier and a real-valued
adder/subtractor, which allows a complex-valued multiplication to be
executed in four cycles.
%
The adder is utilized in only two of the four cycles and in the
remaining two cycles an additional number can be added
to the final result.
%
Hence, the second of the above mentioned computations ($\bV_Y - \bA (\bV_X \bA^H)$)
can be completed with the same rectangular array.
Furthermore, the flow of data through the array and the rate at which
input data must be provided is very regular.
%
An additional advantage of this architecture becomes apparent when a
matrix multiplication is followed by a matrix multiplication with
subtraction as in the compound node example of Fig.~\ref{fig:cn}.
In this case, the result of the first matrix multiplication is stored
in the {\em StateReg} of the PEmult.
The subsequent $\bV_Y - \bA (\bV_X \bA^H)$ operation can immediately
be started since the matrix $\bV_X \bA^H$ is already available and the
matrices $\bA$ and $\bV_Y$ can be shifted through the array from the
``north'' and the ``west'', respectively.

\begin{figure}[h!]
\centering
\includegraphics[width=\columnwidth]{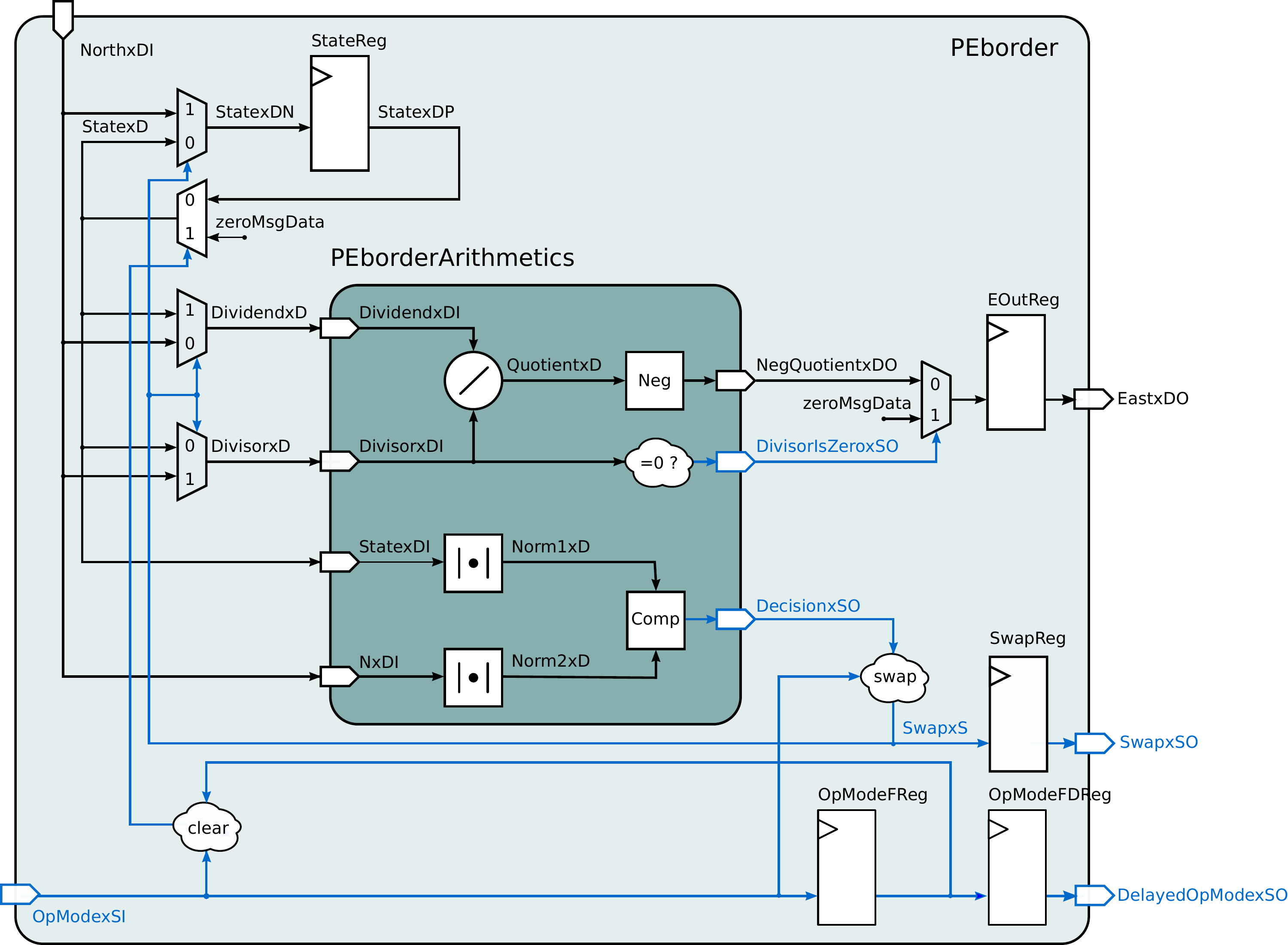}
\caption{{\em PEborder} with different operation modes.}
\label{fig:PEborder}
\end{figure}

The Schur complement for the third operation mode is computed with the
Faddeev algorithm.
The algorithm performs triangularization of the input matrix followed
by Gaussian elimination.~\cite{syst_kalman}.
It is a favorable choice because it avoids direct matrix inversions.
To this end, the rectangular systolic array requires a triangular
extension as it is shown in the yellow box in Fig~\ref{fig:FGP}.
A different kind of PEs capable of computing the absolute value and a
complex division as shown in Fig.\,\ref{fig:PEborder} is required in
addition the PEmult.
During Gaussian elimination, the PEmult performs the necessary swapping
of the matrix rows for the pivoting.

The complex-valued division is carried out using the relationship
$\frac{a+bi}{c+di}=\frac{ac+bd}{c^2+d^2}+\frac{bc-ad}{c^2+d^2}i$ by
employing one sequential divider\footnote{The divider performs a
  sequential radix-2 division in 4 cycles.}, two multipliers and one
adder.

In order to avoid the deployment of three different systolic arrays we
propose an architecture with a single systolic array whose processing
elements support multiple modes of operation as explained in the
following section.

\section{FGP Architecture}

\begin{figure*}[t]
\centering
\includegraphics[width=\textwidth]{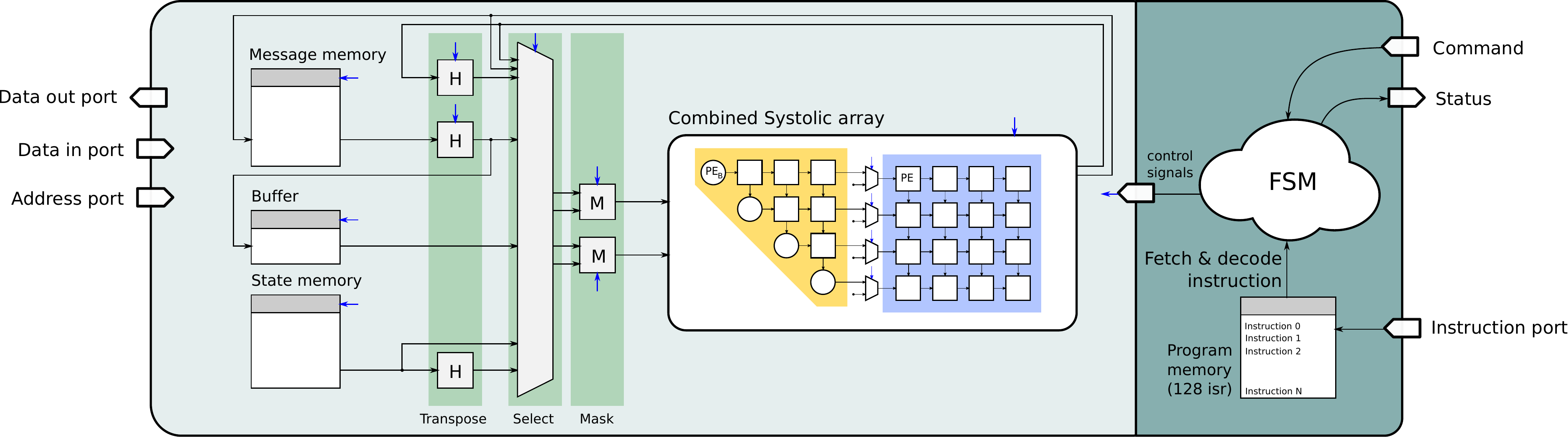}
\caption{FGP architecture with systolic array and memories. The Select
  and Mask units are used to feed data from different sources to the
  array. Processor operation is controlled by a finite state machine
  whose state transitions are triggered from exernal
  commands. Internal Control signals are marked by blue arrows.}
\label{fig:FGP}
\end{figure*}

The FGP architecture is mainly determined by the operations of the GMP
nodes.
The systolic array with the presented PEs is the core part of the FGP
as shown in Fig.\,\ref{fig:FGP}.
Further more, the processor contains a program memory (PM) and
memories for the messages and the state matrix $\bA$.
Storing intermediate results during the execution of an operation is
not required due to the systolic architecture where intermediate
results are stored in the state of the systolic array (i.e., the
result of the matrix multiplication in \textit{accum} mode, which is
used as input to the matrix multiplication in \textit{shift} mode and
as input to the Faddeev algorithm).
%
%

An instruction is fetched from the PM, decoded and forwarded to a
finite state machine which generates the necessary control signals for
the PEs as well as for the Transpose-, Select- and Mask-unit.
The processor operates on a proper instruction set tailored to GMP
called {\em FGP\,Assembler}, comprising six instructions which are
summarized in Table\,\ref{tbl:isr}.

\begin{table}[h!]
\renewcommand{\arraystretch}{1.3}
\caption{FGP Assembler instructions}
\label{tbl:isr}
\centering
    \begin{tabular}{|l|l|}
    \hline
    \bf{Datapath Control Instruction}  & ~                        \\ \hline\hline
    \bf{\texttt{mma}}                         & Matrix multiplication \& accumulate \\ \hline
    \bf{\texttt{mms}}                         & Matrix multiplication \& shift      \\ \hline
    \bf{\texttt{fad}}                         & Faddeev algorithm                 \\ \hline\hline
    \bf{Program Control Instruction} & ~                             \\ \hline\hline
    \bf{\texttt{smm}}                         & Store result of array in memory                      \\ \hline
    \bf{\texttt{loop}}                        & loop over instructions (FG sections)    \\ \hline
    \bf{\texttt{prg}}                         & Program (define multiple programs) \\ \hline
    \end{tabular}
\end{table}

They can be subdivided into instructions which control the program flow and
instructions which control the systolic array.
Each instruction of the latter type corresponds to one computation
type of Section~\ref{sec:datapath}.
The arguments of the instructions are the addresses of the input and
output messages in the memory as well as flags for the Hermitian
transpose and negation.
Since many factor graphs show a repetitive pattern (e.g., RLS) an
instruction for looping over iterations is provided.
In order to host multiple programs in the PM, the {\bf\texttt{prg}}
instruction was introduced to indicate the start addresses of the
different programs.
For example a baseband receiver might store one program for RLS
channel estimation and another one for symbol detection/equalization.

The FGP can be controlled from an external processor via a set of
commands. Each command gets replied by a status message.
Elementary commands are \texttt{load\_program} and
\texttt{start\_program} which load one or multiple programs from
the instruction port into the PM or start a program from the
PM respectively.
In this way the FGP can be easily attached to an existing system as an
accelerator or a co-processor.
The initial input messages need to be loaded into the message memory
via the \textit{Data in} port. After program execution, the results
can be obtained from the message memory through the \textit{Data out} port.
%



\section{Hardware-Software Interaction example for channel estimation}

\begin{figure}[h!]
\centering
\includegraphics[width=0.9\columnwidth]{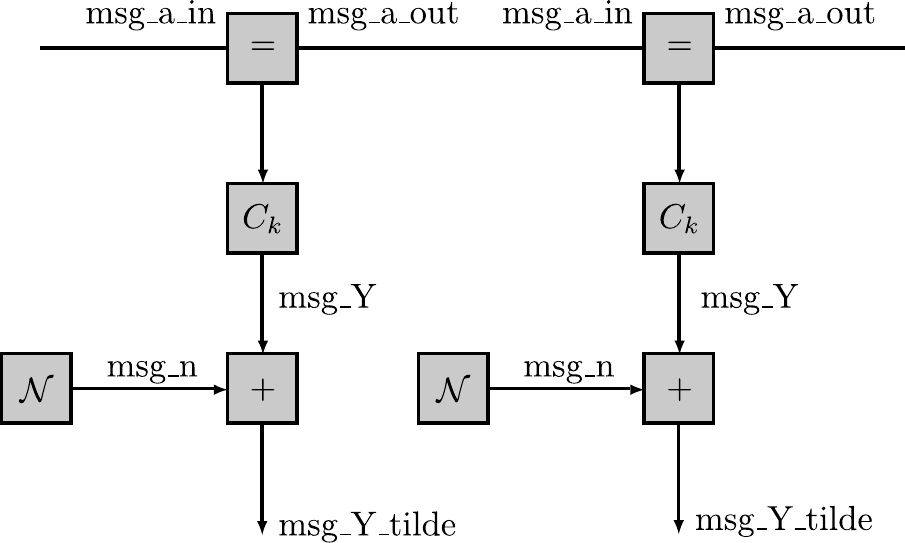}
\caption{Factor graph for channel estimation with two sections.}
\label{fig:RLS_fg}
\end{figure}

The desired GMP algorithm is first written in a high-level language
(Matlab) and then automatically compiled to FGP Assembler code. This
procedure is exemplified for an (Linear Minimum Mean Squared Error)
LMMSE channel estimation example using the FG in
Fig.\,\ref{fig:RLS_fg} and the code listings \ref{lst:FGPM} and
~\ref{lst:FGPA}.
For the sake of simplicity the RLS factor graph contains only two
sections.
In the channel estimation example the messages {\texttt{msg\_Y}} correspond to the
received symbols.

\begin{figure}[h!]
\centering
\includegraphics[width=\columnwidth]{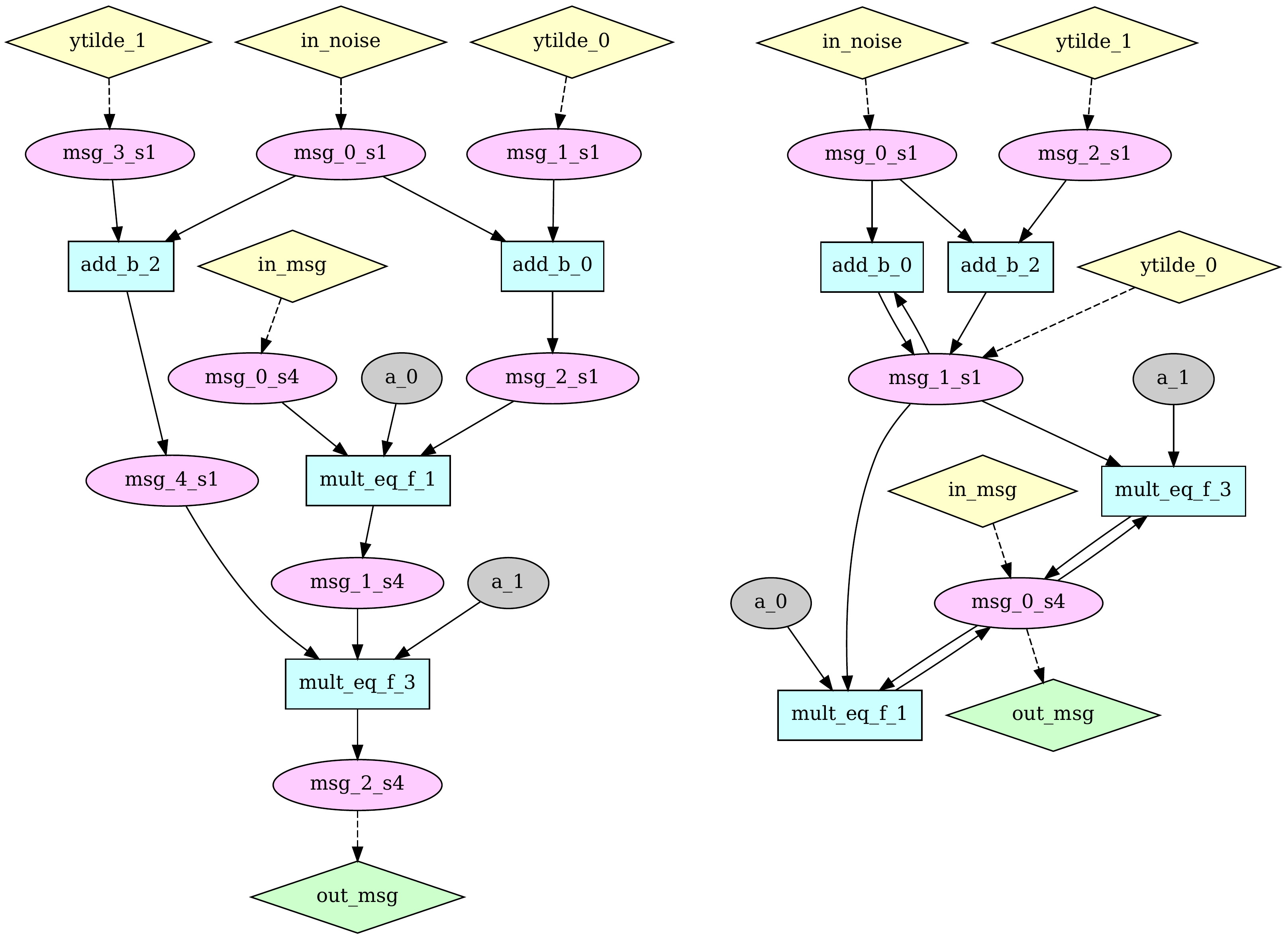}
\caption{Compiler-generated RLS message computation graph with memory
  locations. Left: Unoptimized. Right: Optimized.}
\label{fig:msgupdate}
\end{figure}

A message update schedule (Fig.\,\ref{fig:msgupdate} left) is first derived
from the high level description in Listing~\ref{lst:FGPM}.
\lstset{basicstyle=\footnotesize\ttfamily,
xleftmargin=\parindent,
language=Matlab,
caption={Matlab code for the channel estimation example},
label=lst:FGPM,
frame=l,numbers=left}
\begin{lstlisting}
for i = 1:length(ytilde)
 % observation message
  msg_Ytilde.m = ytilde(i);

  msg_Y = FGP.add_b(msg_Ytilde, msg_n);
  msg_a_out = FGP.mult_eq_f(msg_a_in, msg_Y, A);

  % Update input message for next slice   
  msg_a_in = msg_a_out;
end
\end{lstlisting}

Each message has an identifier assigned, which is mapped to a memory
address later on.
In a second step, the schedule is optimized to reduce the number of
identifiers and hence the size of the message memory
(Fig.\,\ref{fig:msgupdate} right).
Sequentially, for each output message, the set of identifiers assigned
to messages that are no longer needed is considered.
A score is computed for each identifier in the set and the
output message will be remapped to the identifier having the highest
score.
The optimized schedule is then compiled into an assembly language
program (Listing~\ref{lst:FGPA}).
This program is compressed using the loop instruction and converted
into a binary memory image suitable for loading into the processor.
%

        
        
        

        
        

\lstset{basicstyle=\footnotesize\ttfamily,
xleftmargin=\parindent,
language=[x86masm]Assembler,
caption={FGP Assembler code for the channel estimation example},
label=lst:FGPA,
frame=l,numbers=left}
\begin{lstlisting}
prg  1
loop 1 1
mma  1 1 c 0 1 e 0 0 0
mms  0 1 d 0 1 e 1 0 0
smm  1 1 d 0
mma  0 4 d 0 4 c 0 1 0
mms  1 1 d 0 4 c 1 0 0
fad  0 4 d 1
smm  0 4 d 1
\end{lstlisting}

\section{Implementation Results}


As a proof of concept the design was synthesized for the UMC180 CMOS
technology for a state matrix size of 4x4 and compared to a
state-of-the-art DSP in terms of throughput.
As a performance metric the number of message updates on a compound
node ($\mathrm{CN}$) per second was chosen because the compound node
is the GMP node with the highest computational load.
The number of cycles the C66x DSP~\cite{C66x} would take for execution is estimated
using the DSPs fixed-point instruction set.
%
According to \cite{inv}, 768 cycles for the inversion of a complex 4x4
matrix are assumed.
%
%
%
%
The comparison is shown in Table~\ref{tbl:cmp}.

\begin{savenotes}
\begin{table}[h!]
\renewcommand{\arraystretch}{1.3}
\caption{Throughput comparison, FGP vs DSP}
\label{tbl:cmp}
\centering
    \begin{tabular}{|l|l|l|}
        \hline
        \bf{Processor}                                   & FGP (this work)  & TI C66x~\cite{C66x}     \\ \hline\hline
        CMOS technology [nm]                             & 180        & 40           \\  \hline
        Max. freq. [MHz]                             & 130        & 1250         \\ \hline
        cycles for compound-node msg. update          & 260        & 1076        \\ \hline
        Normalized max. throughput\footnote{Technology scaling to 180nm CMOS assuming $t_{pd} \sim 1/s$.}  [$\mathrm{CN}/$s]               & 2.25$\cdot10^6$        & 1.16$\cdot10^6$         \\ \hline
    \end{tabular}
\end{table}
\end{savenotes}

%
The FGP occupies an area of 3.11\,mm$^2$ of which 30\% are memories, 60\%
systolic array and 10\% datapath and control logic.
The maximal clock speed is 130\,MHz.
The throughput of the C66x is computed at a clock frequency of
1.25GHz~\cite{C66x}.
Both the C66x DSP and the FGP operate in fix point number
representation and have 64kbit of memory.

Due to the usage of the Faddeev algorithm on the systolic architecture
and the usage of a dedicated radix-2 divider, the FGP is able to
compute a compound node message update in only 260 cycles.
Computing the Schur complement using the Faddeev algorithm is more
efficient than computing the inverse and the summands in the Schur
complement separately on a conventional DSP instruction set.
This leads to a 2x higher throughput for the FGP as can be seen in
Table\,\ref{tbl:cmp}.

\section{Conclusion}

The FGP is a promising processing architecture built on the theoretical
GMP framework.
It is able to run a wide range of signal processing algorithms.
The use of a highly configurable systolic array architecture with a
dedicated radix-2 divider allows an efficient computation of message
updates on a Gaussian factor graph.
The presented approach does not only allow to use a low-complexity
fixed-point systolic architecture, it also just needs a very simple
compilation procedure instead of a complex tool-chain, since the user
specifies the message update rules in the high-level representation of
the algorithm.
In a simple channel estimation example we have shown how the FGP
achieves a 2x higher throughput over a state-of-the art DSP.
However, neither the presented processor architecture nor it's
applications are fully explored.
Future work will include optimizations on architectural level which
further exploit the characteristics of GMP, as well as the
implementation and performance comparison for other algorithms.

\bibliographystyle{IEEEtran} \bibliography{IEEEabrv,./paper_arxiv}

\flushend

\end{document}